\documentclass[]{aa}
\usepackage{natbib}
\usepackage{amsmath}
\usepackage{amssymb}
\usepackage{graphicx}

\newcommand{\be}{\begin{equation}}
\newcommand{\ee}{\end{equation}}
\newcommand{\mkp}{ }

\begin{document}

\title{Reacceleration of electrons in supernova remnants}

\author{M. Pohl\inst{1,2}\fnmsep\thanks{Corresponding author}
\and
A. Wilhelm\inst{1,2}
\and
I. Telezhinsky\inst{1,2}
}

\institute{DESY, 15738 Zeuthen, Germany 
\and
Institute of Physics and Astronomy, University of Potsdam, 14476 Potsdam, Germany \\
\email{marpohl@uni-potsdam.de}}

\date{Received / Accepted}

\abstract{The radio spectra of many shell-type supernova remnants show deviations from those expected on theoretical grounds.}{In this paper we determine the effect of stochastic reacceleration on the spectra of electrons in the GeV band and at lower energies, and we investigate whether or not reacceleration can explain the observed variation of radio spectral indices.}{ We explicitely calculate the momentum diffusion coefficient for 3 types of turbulence expected downstream of the forward shock: fast-mode waves, small-scale non-resonant modes, and large-scale modes arising from turbulent dynamo activity. Noting that low-energy particles are efficiently coupled to the quasi-thermal plasma, a simplified cosmic-ray transport equation can be formulated and is numerically solved.}{Only fast-mode waves can provide momentum diffusion fast enough to significantly modify the spectra of particles. Using a synchrotron emissivity that accurately reflects a highly turbulent magnetic field, we calculate the radio spectral index and find that soft spectra with index $\alpha\lesssim -0.6$ can be maintained over more than 2 decades in radio frequency, even if the electrons experience reacceleration for only one acceleration time. A spectral hardening is possible but considerably more frequency-dependent. The spectral modification imposed by stochastic reacceleration downstream of the forward shock depends only weakly on the initial spectrum provided by, e.g., diffusive shock acceleration at the shock itself.}{}

\keywords{Turbulence -- Acceleration of particles -- ISM: supernova remnants -- cosmic rays}


\maketitle

\section{Introduction}
The synchrotron spectra observed from shell-type supernova remnants (SNR) are conventionally interpreted as being produced by electrons that have been accelerated at the forward shock and possibly the reverse shock \citep{2008ARA&A..46...89R} through a process known as diffusive shock acceleration \citep{1978MNRAS.182..147B}. The synchrotron spectra may extend to the hard X-ray band \citep{1995Natur.378..255K}, implying acceleration beyond 10~TeV electron energy and a potentially significant inverse-Compton emission component in the TeV band \citep{1996A&A...307L..57P}. Whereas the electron spectrum at very high energies is shaped by energy losses and the structure of the cosmic-ray precursor, GeV-band electrons should not be affected by losses and boundary effects. Their spectrum should reflect the canonical solution $N(E)\propto E^s$ \mkp{with $s=2$} for strong shocks in monoatomic hydrogen gas. A slight softening of the spectra may arise from cosmic-ray feedback on the shock structure \citep{1987PhR...154....1B} and a proper motion of the cosmic-ray scattering centers, such that the compression ratio of the scattering centers is lower than that of the gas.

The GeV-scale spectrum of electrons is probed with measurements of their synchrotron emission in the radio band. \citet{2009BASI...37...45G} has compiled a catalogue of 274 Galactic SNRs that includes information on the spectral index of their integrated radio emission. The values of the radio spectral index display a large scatter around a mean of $\alpha \approx -0.5$ ($S_\nu\propto \nu^\alpha$, \mkp{with $\alpha=(s-1)/2$}), reaching in some cases $\alpha \approx -0.2$ or $\alpha \approx -0.8$. It is easy to see possible systematic uncertainties in the measurements. Contributions from a pulsar-wind nebula may harden the spectrum, but at least in cases with high-quality data over a wide frequency range a lack of strong curvature in the spectrum would argue against that possibility. Likewise, confusion with H-$II$ regions should be identifiable. Improper background subtraction should not be a problem for bright SNR like Cas~A, which has a spectral index of $\alpha \approx -0.77$.

It was realized early that stochastic acceleration might be a second relevant acceleration process \citep{1983SSRv...36...57D}. Its efficiency was perceived to be low, and so few attempts have been made to explain the entire electron acceleration in SNR on this basis \citep{2008ApJ...683L.163L}. If outward-moving Alfv\'en or fast-mode waves are resonantly excited in the shock precursor, their passage through the shock leads to a mixture of forward and backward moving waves \citep{1999A&A...343..303V}. The Alfv\'en speed is small, though, and so is the momentum-diffusion coefficient calculated for scattering on them \citep[e.g.][]{2002cra..book.....S}. Fast-mode waves are a possible alternative, because their phase velocity is large in the downstream region \citep{2008ApJ...683L.163L}. Non-resonant wave production typically yields linear waves with negligible phase velocity \citep[e.g.][]{2004MNRAS.353..550B} and hence negligible efficacy for stochastic acceleration.

Here we re-examine the role of stochastic acceleration in SNR. It has been realized in recent years, that non-resonant small-scale instabilities operating upstream in their non-linear phase impose substantial plasma turbulence that will foster second-order Fermi acceleration \citep{2009ApJ...706...38S}. Secondary instabilities arise, for example by shock rippling, which leads to turbulent magnetic-field amplification downstream of the shock \citep{2007ApJ...663L..41G,2011ApJ...726...62M,2011ApJ...734...77I,2012ApJ...747...98G,2013ApJ...770...84F}, along with turbulent motions that should eventually be in energy equipartition with the turbulent magnetic field \citep{2014MNRAS.439.3490M}. Both on small and on large scales we therefore expect some second-order Fermi acceleration to operate behind the outer shocks of SNRs. The inevitable decay of the turbulence will not only impose a spatial dependence on the acceleration rate, it will also impact on the synchrotron emissivity \citep{2005ApJ...626L.101P}. In fact, numerical modeling of particle motion in generalized MHD turbulence suggests that for parameters typical of young SNRs a GeV-scale particle can experience acceleration on a time scale of $\sim 100$~years \citep[their experiment 9]{2014ApJ...784..131F}. Stochastic acceleration may thus act as a secondary re-acceleration process downstream of SNR shocks that slightly modifies the particle spectrum produced at the shock by diffusive shock acceleration. This is in contrast to some models of flares that posit an initial stochastic acceleration followed by a second stage of shock acceleration \citep{2012SSRv..173..535P}.

In this paper we attempt an estimate of the re-acceleration rate for three types of turbulence: fast-mode waves as already discussed by \citet{2008ApJ...683L.163L}, Bell's non-resonant instability, and large-scale MHD turbulence arising from shock rippling through dynamo processes. Having established the efficiency, energy dependence, and spatial decay scale of the momentum diffusion coefficient, we compute its effect on the differential number density of electrons between the forward shock and the contact discontinuity. We conclude with a discussion of the expected radio spectra of SNRs.

\section{Estimating the rate of diffusive reacceleration}\label{diffacc}
Before we discuss reacceleration rates in detail, it is important to recall that the downstream region of the forward shocks of young SNRs is not a low-$\beta$ environment. The flow is subsonic by definition, whereas the Alfv\'en speed increases only with the shock compression. It has become popular to posit a large Alfv\'en velocity of a few hundred kilometers per second in the upstream region, which leads to softer particle spectra that fit better to observations. This benefit comes at the expense of reducing the Alfv\'enic Mach number of the forward shock to $M_\mathrm{A}\lesssim 10$, rendering the magnetic field dynamically important \citep{2008ApJ...679L.139C} and an efficient energy sink.

\subsection{Fast-mode waves}
The fastest waves in the downstream region should be fast-mode waves whose phase velocity is the sound speed, $v_\mathrm{fm}= c_\mathrm{s}\simeq 1000$~km/s. Particles can interact with fast-mode waves through transit-time damping (TTD), a process that does not have specific resonance scales. One consequence is that thermal particles will efficiently damp all waves except those that propagate parallel or perpendicular to the local magnetic field \citep{1998ApJ...500..978Q}. The damping rate of waves with propagation angle $\theta_k$ to the large-scale magnetic field is of the order \citep{2008ApJ...683L.163L}
\begin{align}
&\Lambda_\mathrm{d,fm}\approx \frac{2\pi\,c_\mathrm{s}\,\sin^2\theta_k}{\lambda\,\vert\cos\theta_k\vert}\ \nonumber \\
&\cdot \left(
\exp\left[-\frac{1}{\cos^2\theta_k}\right]+\sqrt{\frac{m_e\,T_e}{m_p\,T_p}}\,\exp\left[-\frac{m_e\,T_p}{m_p\,T_e\,\cos^2\theta_k}\right]\right) \ .
\label{eq:1.1}
\end{align}
In the immediate postshock region of young SNRs the electron temperature, $T_e$, is considerably lower than that of protons. In fact, $T_e/T_p \propto v_\mathrm{sh}^{-2}$ and $T_e \approx 0.01\,T_p$ for shock speeds, $v_\mathrm{sh}$, around $4,000$~km/s \citep{2007ApJ...654L..69G,2008ApJ...689.1089V}. The prefactor of the second exponential in Equation~\ref{eq:1.1} is therefore of the order $2\cdot 10^{-3}$, and the argument of the second exponential is approximately $1/(20\,\cos^2\theta_k)$
An estimate of the time scale of cascading is \citep{2002PhRvL..88x5001C}
\be
\tau_\mathrm{c,fm}\approx \ \frac{c_\mathrm{s}\,\sqrt{\lambda\,\lambda_\mathrm{max}}}{2\pi\,V^2}\ ,
\label{eq:1.2}
\ee
where $V$ is the amplitude of velocity fluctuations at the injection scale, $\lambda_\mathrm{max}$. We see that the entire spectrum of fast-mode waves will be anisotropic. Setting $\Lambda_\mathrm{d,fm}\,\tau_\mathrm{c,fm}=1$ gives the angle-dependent wavelength, $\lambda_\mathrm{c}$, down to which the fast-mode turbulence can cascade. 

The fast-mode turbulence can be expected to follow a 3D-spectrum \citep{2002PhRvL..88x5001C}
\be
W(k,\Omega_k)=W_0\,k^{-3.5}\,\Theta\left(k-\frac{2\pi}{\lambda_\mathrm{max}}\right)\,\Theta\left(\frac{2\pi}{\lambda_\mathrm{c}}-k\right) \ , 
\label{eq:1.4}
\ee
where \mkp{$\Theta$ is a step function and}
\be
W_0=\frac{\delta B_\mathrm{fm}^2\,\sqrt{2\pi}}{16\pi\,\sqrt{\lambda_\mathrm{max}}}
=\frac{\rho\,V^2}{\beta}\,\frac{\sqrt{2\pi}}{16\pi\,\sqrt{\lambda_\mathrm{max}}}
=\frac{U_\mathrm{fm}\,\sqrt{2\pi}}{8\pi\,\beta\,\sqrt{\lambda_\mathrm{max}}}
\label{eq:1.5}
\ee
with $U_\mathrm{fm}$ denoting the kinetic energy density in fast-mode waves and $\lambda_\mathrm{max}$ is their driving scale. This spectrum refers to the magnetic fluctuations only, which are known to be weaker in energy density than the velocity fluctuations, and so the plasma beta appears in the denominator.

We calculate the momentum diffusion coefficient for an isotropic distribution of electrons as \citep{2014ApJ...791...71L}
\be
D_p=\int d\mu\ \frac{p_\perp^2\,v_\perp^2}{8\,B_0^2}\,\int dk\ k^2 \oint d\Omega_k\ k_\parallel^2\,
W(k,\Omega_k)\,R(k,\Omega_k)\ ,
\label{eq:1.6}
\ee
where $\perp$ and $\parallel$ refer to projections perpendicular and parallel to the local mean magnetic field, $B_0$, and $\mu=\cos\theta$ reflects the pitch angle relative to it. The resonance function $R(k,\Omega_k)$ includes the effects of orbit perturbations and can be written as \citep{2008ApJ...673..942Y}
\begin{align}
&R(k,\Omega_k)=\frac{1}{\sqrt{\pi}\,\vert \epsilon\,k_\parallel\,v_\perp\vert}\,
\exp\left[-\frac{(k_\parallel\,v_\parallel-k\,c_s)^2}{\epsilon^2\,k_\parallel^2\,v_\perp^2}\right]\nonumber \\
&=\frac{1}{\sqrt{\pi}\,\epsilon\,k\,c\,\vert \mu_k\vert\,\sqrt{1-\mu^2}}\,
\exp\left[-\frac{(\mu-\frac{c_s}{\mu_k\,c})^2}{\epsilon^2\,(1-\mu^2)}\right]
\ ,
\label{eq:1.7}
\end{align}
where $\mu_k=k_\parallel/k$ and $\epsilon \simeq (\langle\delta B_\parallel^2\rangle/B^2)^{1/4}$ involves the large-scale fluctuations in the magnetic field parallel to the mean field and should be of the order unity in the highly perturbed environment immediately downstream of a SNR forward shock.

Inserting Equation~\ref{eq:1.4} and \ref{eq:1.7} into \ref{eq:1.6} we find
\begin{align}
D_p=&\frac{\sqrt{\pi}\,p^2\,c\,U_\mathrm{fm}}{8\,\epsilon\,B_0^2\,\beta\,\sqrt{\lambda_\mathrm{max}}}\,
\int d\mu\ \left(1-\mu^2\right)^{3/2}\nonumber \\
&\cdot \int d\mu_k\ \frac{\vert \mu_k\vert}{\sqrt{\lambda_c}}\,
\exp\left[-\frac{(\mu-\frac{c_s}{\mu_k\,c})^2}{\epsilon^2\,(1-\mu^2)}\right]
\ .
\label{eq:1.8}
\end{align}
We solve Equation~\ref{eq:1.8} separately for parallel and perpendicular waves in the appendices~\ref{a1} and \ref{a2}, respectively. We cannot confirm the finding of 
\citet{2008ApJ...683L.163L} who argue that the parallel modes are the most efficient accelerators of electrons. In fact, the perpendicular modes give a momentum diffusion coefficient that is somewhat larger than that of parallel modes.

From Equation~\ref{eq:a8} we find the acceleration time as
\begin{align}
&\tau_\mathrm{acc}\simeq \frac{p^2}{D_p}\approx
\frac{40\,\sqrt{\pi}\,\sqrt{\lambda_\mathrm{max}\,\lambda_\mathrm{min}}}{c\,\mu_{k,c}^2}\,\frac{U_\mathrm{th}}{U_\mathrm{fm}}\nonumber \\
&\simeq (2\cdot 10^7\ \mathrm{s})\,
\left(\frac{\sqrt{\lambda_\mathrm{max}\,\lambda_\mathrm{min}}}{10^{13}\ \mathrm{cm}}\right)\,
\left(\frac{U_\mathrm{th}}{10\ U_\mathrm{fm}}\right)\left(\frac{\mu_{k,c}}{0.1}\right)^{-2}\ .
\label{eq:1.9}
\end{align}
This estimate for the time scale of stochastic acceleration is independent of momentum. It does rely on the isotropy of the particle distribution function, though. Calculating the isotropization time scale is beyond the scope of this paper, but a remark may be in order. Usually, isotropization becomes slower at high particle energies, and therefore Equation~\ref{eq:1.9} should be realistic only for low-energy particles with energy $E\lesssim 1\ \mathrm{GeV}$.

Reaccelerating energetic particles is another damping process for fast-mode turbulence that we have not yet considered. Let us now consider a not too narrow range in $k$ of the wave spectrum around wavelength $\lambda$. The wave energy density at that wavelength, and the acceleration time provided by it, are
\be
U_\mathrm{fm}(\lambda)\simeq U_\mathrm{fm}\,\sqrt{\frac{\lambda}{\lambda_\mathrm{max}}}\qquad
\tau_\mathrm{acc}(\lambda)\simeq \tau_\mathrm{acc}\,\sqrt{\frac{\lambda_\mathrm{min}}{\lambda}}\ .
\label{eq:1.9c}
\ee
Taking the acceleration time independent of energy as in Equation~\ref{eq:1.9}, we find the energy transfer rate from turbulence to particles as
\be
\dot E_\mathrm{tr}(\lambda) \simeq \int dE\ \frac{E\,N(E)}{\tau_\mathrm{acc}(\lambda)}=\frac{U_\mathrm{cr,acc}}{\tau_\mathrm{acc}(\lambda)}\ ,
\label{eq:1.9a}
\ee
where $U_\mathrm{cr,acc}$ denotes the energy density in cosmic rays that experience acceleration. We noted before that very-high-energy particles will probably not isotropize fast enough for our estimate of the acceleration time to be valid. The energy transfer, $\dot E_\mathrm{tr}(\lambda)$, reduces the energy density in fast-mode waves, $U_\mathrm{fm}(\lambda)$ on a time scale
\begin{align}
\tau_\mathrm{d,cr}(\lambda)&\simeq \frac{U_\mathrm{fm}(\lambda)}{\dot E_\mathrm{tr}(\lambda)}\simeq
\frac{U_\mathrm{fm}(\lambda)}{U_\mathrm{cr,acc}}\,\tau_\mathrm{acc}(\lambda) \nonumber \\
&\simeq \frac{U_\mathrm{fm}}{U_\mathrm{cr,acc}}\,\tau_\mathrm{acc}\,\frac{\lambda}{\sqrt{\lambda_\mathrm{max}\,\lambda_\mathrm{min}}}\nonumber \\
&\simeq \frac{U_\mathrm{th}}{U_\mathrm{cr,acc}}\,\frac{40\,\sqrt{\pi}\,\lambda}{c\,\mu_{k,c}^2}\ .
\label{eq:1.9b}
\end{align}
The cosmic-ray induced damping of the waves must be slower than cascading, otherwise the fast-mode cascade would terminate. Comparing Eqs.~\ref{eq:1.9b} and~\ref{eq:1.2} we find
\be
\lambda\gtrsim \lambda_\mathrm{max}\,\left(\frac{c}{9\cdot 10^4\ c_\mathrm{s}}\right)^2\,
\left(\frac{U_\mathrm{cr,acc}}{U_\mathrm{fm}}\right)^2\,
\left(\frac{\mu_{k,c}}{0.1}\right)^4\ .
\label{eq:1.9d}
\ee
Combined, Equations~\ref{eq:1.9b} and~\ref{eq:1.9d} illustrate conditions that must be met
for stochastic reacceleration to be operational in the postshock region of SNR.
\begin{enumerate}
\item There should not be less energy density in fast-mode waves than in the part of the cosmic-ray spectrum in which re-acceleration is efficient. 

\item Unless we have \emph{much} more energy in fast-mode turbulence than in energetic particles undergoing stochastic acceleration, cascading is quenched by TTD interaction with cosmic rays. Consequently, our estimate of the acceleration time scale in Equation~\ref{eq:1.9} is optimistic.

\item Inserting the wavelength limit (Eq.~\ref{eq:1.9d}) as $\lambda_\mathrm{min}$ into the expression (Eq.~\ref{eq:1.9}) we find as more realistic, revised estimate of the acceleration timescale
\begin{align}
&\tau_\mathrm{acc,rev}\simeq \frac{\lambda_\mathrm{max}}{4\pi\,c_\mathrm{s}}\,
\frac{U_\mathrm{th}\,U_\mathrm{cr,acc}}{U_\mathrm{fm}^2} \\
&\simeq (8\cdot 10^7\ \mathrm{s})\,
\left(\frac{\lambda_\mathrm{max}}{10^{16}\ \mathrm{cm}}\right)\,
\left(\frac{c_\mathrm{s}}{10^3\ \mathrm{km/s}}\right)^{-1}\,
\frac{U_\mathrm{th}\,U_\mathrm{cr,acc}}{10\,U_\mathrm{fm}^2} \ . \nonumber 
\label{eq:1.9e}
\end{align} 
The true acceleration time is thus of the order of a few years, and stochastic reacceleration will operate for only a few acceleration times. 

\item As the turbulence is driven at the shock and then advects downstream, we must expect that strong fast-mode turbulence exists only in a thin layer of a few 
$\lambda_\mathrm{max}$ in thickness.
\end{enumerate}
\subsection{Small-scale non-resonant modes}
Current-driven instabilities can lead to aperiodic small-scale turbulence \citep{1984JGR....89.2673W}, that include parallel \citep{2004MNRAS.353..550B} and oblique modes \citep{2014ApJ...780..175M}. In conditions typical for the cosmic-ray precursors of the forward shocks of young SNR, the instability may operate for only a few growth times \citep{2008ApJ...684.1174N}, unless the remnant expands into a high-density environment,
in which case, however, one has to consider ion-neutral collisions, which
can reduce the growth of the instability \citep{2007A&A...475..435R}.

The saturation level of small-scale non-resonant modes is not well known. Estimates based on the shrinking of the driving particles' Larmor radius lead to high saturation levels of a few hundred $\mu G$ \citep{2008ApJ...678..255Z,2006A&A...453..181P}. A significant backreaction, however, is a reduction in streaming velocity and hence a diminishing of the streaming anisotropy, that leads to a lower saturation level and only moderately amplified magnetic field \citep{2009MNRAS.397.1402L}. Another non-linear side effect is turbulent motion \citep{2009ApJ...706...38S} which will add to the large-scale MHD turbulence discussed in the next subsection.

Relevant for us is that the modes have a very small real frequency, at least in the linear stage, and so we can describe them with reasonable accuracy in the magnetostatic approximation. They are excited on scales well below the Larmor radius of the driving particles. If we place the high-energy cut off in the particle spectra spectrum at 100~TeV and assume a magnetic-field strength of
$10\ \mu$G, then the largest scale would be $\sim 10^{16}$~cm. Particles of lower energies excite modes of smaller wavelength and, depending on the cosmic-ray spectrum, may be actually more efficient in driving small-scale turbulence than are the few particles at the highest energy that carry very little current. Gamma-ray observation of young SNR indicate relatively soft particle spectra, and so a more realistic estimate of the length scale of maximum intensity would be $L_\mathrm{max}\approx (10^{14}-10^{15})$~cm. What turbulence spectrum is eventually established is a difficult question. Behind the shock little, if any, driving will transpire, and the spectrum is determined by cascading and damping until the energy reservoir at $L_\mathrm{max}$ is drained.

\citet{2009A&A...507..589S} have developed a second-order non-linear theory of wave-particle interaction that for an assumed spectrum
\be
I_k \propto \frac{|k\,L_\mathrm{max}|^q}{\left[1+(k\,L_\mathrm{max})^2\right]^{(s+q)/2}}
\label{eq:2.1}
\ee
of strong magnetosonic slab turbulence yields for the momentum diffusion coefficient of relativistic particles with small Larmor radius, $r_\mathrm{L}$, \citep{2012PhPl...19j2901S}
\be
D_p \simeq D(s)\,p^2\,\frac{c}{L_\mathrm{max}}\,\left(\frac{v_\mathrm{A}}{c}\right)^2\,
\,\left(\frac{r_\mathrm{L}}{L_\mathrm{max}}\right)^{s-2}\,\left(\frac{\delta B}{B_0}\right)^{1+s}
\label{eq:2.2}
\ee
where
\be
D(s)=\frac{\sqrt{\pi}}{3\,s}\,\frac{\Gamma\left(\frac{s}{2}\right)}{\Gamma\left(\frac{s-1}{2}\right)}\ ,
\ee
$\delta B$ is the turbulent field amplitude integrated over the entire turbulence spectrum~\ref{eq:2.1}, and $v_A$ is the Alfv\'en velocity. 
The diffusion coefficient would fall off steeply once \mkp{$r_\mathrm{L}\gtrsim r_\mathrm{L,max}$ where $r_\mathrm{L,max} = L_\mathrm{max} \,B_0\,/\,\delta B$}. The expression will certainly become inaccurate in the presence of other turbulence on larger scales, but it may still serve as an approximation. 

The cascading behavior is likewise not well known. A generic Kolmogorov-type estimate \citep{2003A&A...403....1P} would suggest that the intensity at the largest scales decays on a timescale
\be
\tau_d\approx \frac{L_\mathrm{max}}{v_A}\,\frac{B_0}{\delta B}
\simeq (30\ \mathrm{yr})\,\left(\frac{L_\mathrm{max}}{10^{15}\ \mathrm{cm}}\right)\,
\left(\frac{v_A}{10\ \mathrm{km/s}}\right)^{-1}\,\frac{B_0}{\delta B}\ .
\label{eq:2.3}
\ee
The ratio of available time (Eq.~\ref{eq:2.3}) to the acceleration time $p^2/D_p$ \mkp{can be expressed with a step function $\Theta$ to mark the range of applicability of Equation~\ref{eq:2.2}},
\be
\frac{\tau_d\,D_p}{p^2}\simeq D(s)\,\frac{v_\mathrm{A}}{c}\,
\left(\frac{r_\mathrm{L}}{r_\mathrm{L,max}}\right)^{s-2}\,\left(\frac{\delta B}{B_0}\right)^2 \,
\Theta\left(r_\mathrm{L,max} -r_\mathrm{L} \right)\ .\label{eq:2.4}
\ee
It is typically small, unless $v_A$ is very large in the downstream region. 

\subsection{Large-scale MHD turbulence}
The acceleration provided by moving magnetic-field structures is essentially a classical Fermi process. If scatterers move with random velocity $v_s$ and the frequency of collision with these structures is $\omega$, then the momentum diffusion coefficient is \citep[e.g.][]{2013ApJ...777..128L}
\be
D_p\approx p^2\,\omega\,\left(\frac{v_S}{c}\right)^2
\label{eq:3.1}
\ee
\citet{2013ApJ...770...84F} estimates that the magnetic-field amplification factor is determined by the Field length \citep{2006ApJ...652.1331I}, $L_\mathrm{F}$, and the radius of curvature, $R_\mathrm{C}$, of the shock ripples,
\be
\frac{B}{B_0}\approx M_\mathrm{A}\,\sqrt{2\,\frac{L_\mathrm{F}+R_\mathrm{C}}{\theta\,L_\mathrm{F}}} \ ,
\label{eq:3.2}
\ee
where $M_\mathrm{A}$ is the Alfv\'enic Mach number and $\theta$ is the typical angle between the local and the average shock normal. The radius of curvature, $R_\mathrm{C}$, can be related to the length scales of upstream density fluctuations, $L_\rho$, as 
\be
R_\mathrm{C}\approx \frac{L_\rho}{2\,\theta}\qquad \Rightarrow \frac{B}{B_0}\approx
\frac{M_\mathrm{A}}{\theta}\,\sqrt{\frac{L_\rho}{L_\mathrm{F}}}\ .
\label{eq:3.3}
\ee
Inserting numbers appropriate for young SNR one easily finds amplification factors of a thousand or more, because the size of upstream clouds,  $L_\rho$, is typically larger than the thickness of their interface to the dilute medium, $L_\mathrm{F}$. \citet{2012ApJ...744...71I} have performed 3D-MHD simulations with $M_\mathrm{A}\simeq 250$ and find amplification factors considerably less than suggested by Equation~\ref{eq:3.3}: The maximum field strength is about a factor 200 higher than the initial field, and the average field strength is amplified by approximately a factor 5. Similar results have been obtained in 2D-MHD simulations \citep{2011ApJ...726...62M,2014MNRAS.439.3490M}. 

Turbulence spectra are usually difficult to extract on account of the limited spectral range of MHD simulations. Published simulations agree that velocity perturbations approximately follow
a Kolmogorov scaling whereas the spectrum of magnetic perturbations is considerably flatter than that, possibly because the dynamo process has not saturated on large scales. The correlation properties between magnetic and velocity perturbations in the simulations of \citep{2014MNRAS.439.3490M} suggest that we may treat the turbulence structures as magnetic clouds of amplitude $B_k$ moving with random velocity $v_k$. The scattering rate is then determined by the time needed to propagate between clouds, either ballistically or through diffusion with mean free path $\lambda_\mathrm{mfp}$, 
\be
\omega_k\approx \eta_k\,\frac{c\,k}{2\pi}\,\frac{1}{1+\frac{6\pi}{\lambda_\mathrm{mfp}\,k}}\ ,
\label{eq:3.4}
\ee
where $\eta_k$ is the efficiency of the process that is of the order unity only if the particles are indeed reflected upon passage through the magnetic structure, thus requiring 
\be
B_k \gg \frac{B_\mathrm{rms}\,r_\mathrm{L}}{\lambda}\ ,
\label{eq:3.5}
\ee
where the Larmor radius, $r_\mathrm{L}$, is to calculated using $B_\mathrm{rms}$. Hence, our estimate for $\eta_k$ is
\be
\eta_k\approx \left[1+\left(\frac{B_\mathrm{rms}\,r_\mathrm{L}}{B_k\,\lambda}\right)^2\right]^{-1}=\left[1+\left(\frac{B_\mathrm{rms}\,r_\mathrm{L}\,k}{2\pi\,B_k}\right)^2\right]^{-1}\ .
\label{eq:3.6}
\ee
The momentum diffusion coefficient (Equation~\ref{eq:3.1}) is then determined through convolution with the turbulence spectrum,
\be
D_p\approx p^2\,\int d\ln k\quad \omega_k\,\left(\frac{v_k}{c}\right)^2\ .
\label{eq:3.7}
\ee
The single-cloud values of magnetic-field strength and velocity need to be replaced with integrals over the Fourier power spectrum, and hence we replace $v_k^2 =k\,\vert F(v)\vert^2$.
Using Kolmogorov scaling for both velocity and magnetic perturbations,
\be
v_k^2 =\frac{2\,v_\mathrm{rms}^2}{3}\,\left(\frac{k}{k_\mathrm{min}}\right)^{-\frac{2}{3}}\quad 
B_k^2 =\frac{2\,B_\mathrm{rms}^2}{3}\,\left(\frac{k}{k_\mathrm{min}}\right)^{-\frac{2}{3}} \ ,
\label{eq:3.8}
\ee
one finds that the integrand in Equation~\ref{eq:3.7} has a sharp maximum at 
\be
k_\mathrm{c}\approx k_\mathrm{min}^\frac{1}{4}\,r_\mathrm{L}^\frac{-3}{4}\ .
\label{eq:3.9}
\ee
The decay of the turbulence therefore affects the momentum diffusion coefficient in an easily tractable way: We need only consider the decay length at $k_\mathrm{max}$.
The integral~\ref{eq:3.7} yields
\be
D_p\approx \frac{2}{27\pi\,c}\,p^2\,v_\mathrm{rms}^2\,\frac{\lambda_\mathrm{mfp}\,k_\mathrm{min}}{r_\mathrm{L}}\ .
\label{eq:3.10}
\ee
If the spatial diffusion follows Bohm scaling ($\lambda_\mathrm{mfp}\approx r_\mathrm{L}$), then the acceleration time is independent of momentum.

Simulations suggest that the velocity fluctuations reach a few per cent of the shock speed, $v_\mathrm{rms}\simeq v_\mathrm{sh}/20$ \citep{2014MNRAS.439.3490M}. Inserting that in Equation~\ref{eq:3.10} we find for the acceleration time
\begin{align}
\tau_\mathrm{acc}&\simeq \frac{p^2}{D_p} \nonumber \\
&\approx (2.5\cdot 10^4\ \mathrm{yr})\,
\left(\frac{0.03\,c}{v_\mathrm{sh}}\right)^{2}\,
\left(\frac{10^{-16}\ \mathrm{cm^{-1}}}{k_\mathrm{min}}\right)\,
\frac{r_\mathrm{L}}{\lambda_\mathrm{mfp}}\ .
\label{eq:3.11}
\end{align}
This is much longer than the evolutionary time scales of SNR, unless $k_\mathrm{min}$ is very small. Type-Ia supernovae expand into the interstellar medium, in which a significant part of the high-density clouds arise from the thermal instability, and so the Field length, $L_\mathrm{F}\approx 10^{16}$~cm, sets the scale for $k_\mathrm{min}$, thus rendering re-acceleration ignorable. Core-collapse supernovae, on the other hand, expand into the wind zone of their progenitors, which is essentially a collection of clumps of material that are radiatively driven outward \citep{1980ApJ...241..300L} that can survive out to at least 1000 stellar radii \citep{2002A&A...381.1015R}. Spectroscopic evidence for clumping has been presented in \citep{1999ApJ...514..909L,2008AJ....136..548L,2010A&A...521L..55P} with clumps sizes ranging up to approximately one stellar radius \citep{2007A&A...476.1331O}. In that case $k_\mathrm{min}\approx 10^{-13}\ \mathrm{cm^{-1}}$ and the acceleration time is only 20 years or so. The time available is limited by the forward shock of the proto-SNR leaving the region of significant clumping, and we find that clumps need to survive out to more than $10^4$ stellar radii to permit a significant spectral distortion.

\section{Calculation of electron spectra}
The spatial transport of relativistic electrons behind the forward shock of SNR is provided by both advection and diffusion. We concern ourselves with GeV-scale electrons whose spatial-diffusion coefficient is most likely very small inside SNR. If we scale diffusion to the Bohm limit in a $100\,\mu$G magnetic field, 
\be
D_r(p)=\eta\,D_\mathrm{Bohm}(p)=\eta\,\left(3\cdot 10^{20}\ \mathrm{cm^2/s}\right)\,\left(\frac{p}{\mathrm{GeV}/c}\right)\ ,
\label{eq:1}
\ee
then the typical displacement of an electron in a time period $\delta t$ is
\be
\delta z\Big\vert_\mathrm{diff}\simeq \sqrt{D_r(p)\cdot \delta t}=\left(10^{14}\ \mathrm{cm}\right)\,\sqrt{\eta\,\frac{\delta t}{\mathrm{yr}}\,\frac{p}{\mathrm{GeV}/c}} \ .
\label{eq:2}
\ee
In the same time period, advection in the downstream flow with typical speed $v_\mathrm{adv}\simeq v_8\,(1000\ \mathrm{km/s})$ yields a displacement
\be
\delta z\Big\vert_\mathrm{adv}\simeq v_\mathrm{adv}\cdot \delta t=\left(3\cdot 10^{15}\ \mathrm{cm}\right)\,v_8\,\left(\frac{\delta t}{\mathrm{yr}}\right) \ .
\label{eq:3}
\ee
For very small time periods and displacements from the forward shock, 
\begin{align}
\delta t&\lesssim (3\cdot 10^4\ \mathrm{s})\,\frac{\eta}{v_8^2}\,\left(\frac{p}{\mathrm{GeV}/c}\right)\nonumber \\
\delta z &\lesssim (3\cdot 10^{12}\ \mathrm{cm})\,\frac{\eta}{v_8}\,\left(\frac{p}{\mathrm{GeV}/c}\right)\ 
\label{eq:4}
\end{align}
diffusion is faster than advection, and particles can return to the shock for further acceleration. Further downstream, beyond a distance $\sim D_r(p)/v_\mathrm{adv}$ behind the shock, GeV-scale electrons will on average not return to the shock and their transport is predominantly provided by advection, which permits a simplified treatment of particle transport. 

The following approximations are valid with reasonably good accuracy:
\begin{itemize}
\item The spatial transport is predominantly radial and hence a 1-D problem with spherical symmetry. 
\item The differential density of electrons, $N(r,p,t)$, follows a continuity equation that can be written in the local shock rest frame, i.e. with spatial coordinate $z=r_\mathrm{sh}(t)-r$.
\item If advection is the only transport process, then electrons will move on a characteristic in $z$-$t$ space. For simplicity we shall assume a constant advection speed, for which the characteristic is given by Equation~\ref{eq:3}. 
\item If stochastic acceleration processes operate in a thin layer behind the forward shock, they can impact electrons only for a time period much shorter than the age of the SNR. Then, adiabatic losses, curvature of the forward shock, and expansion, essentially all evolutionary effects that operate on the dynamical time scale of the remnant, are ignorable, if we solve for the radial distribution of electrons, $4\pi\,r^2\,N(r,p,t)$, rather than the space density.
\end{itemize}
\begin{figure}[ht]
 \resizebox{\hsize}{!}{\includegraphics[angle=0]{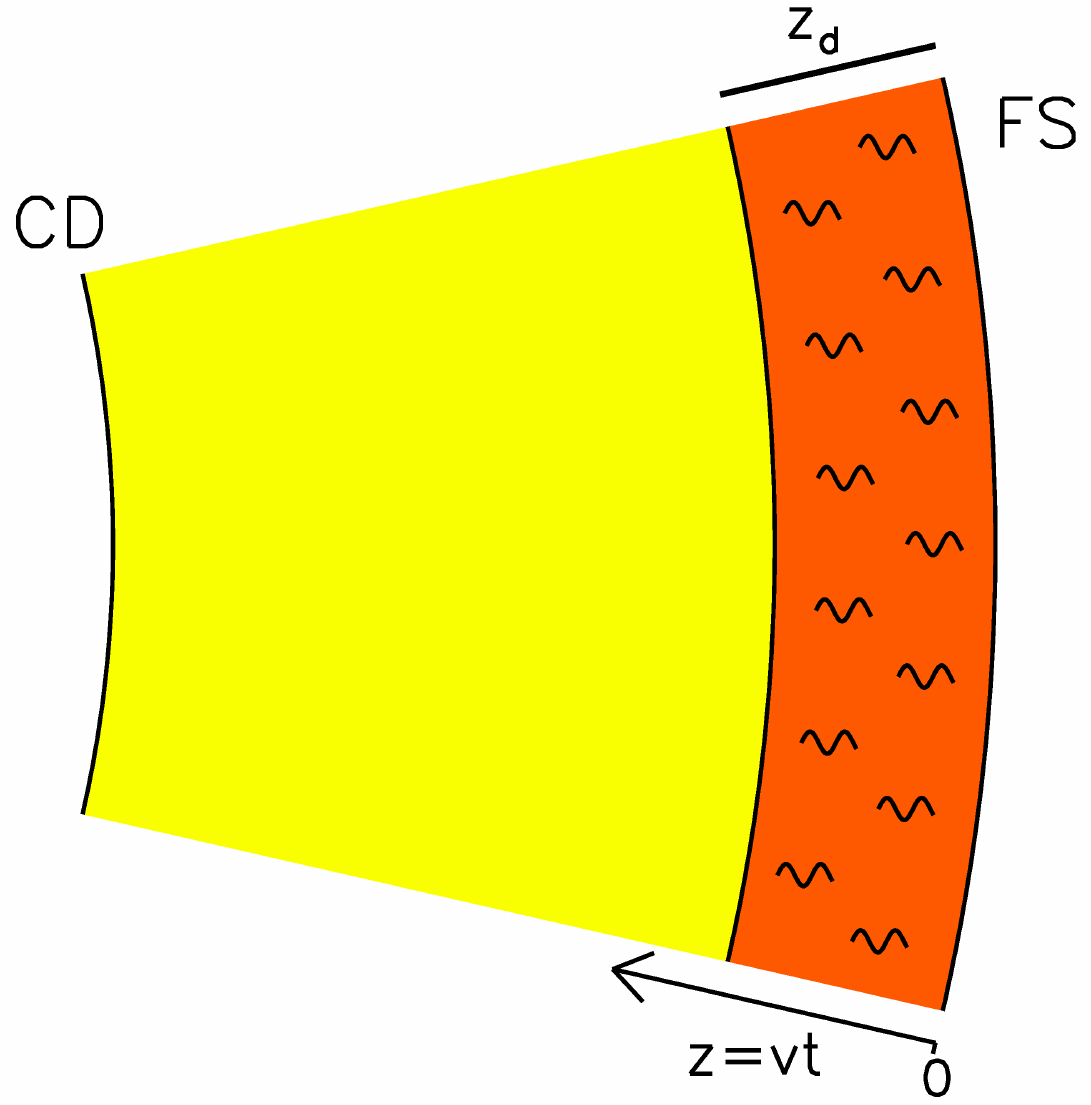}}
 \caption{Schematic representation of the scenario. In a region of thickness $z_d$ behind the forward shock (FS) turbulence can re-accelerate electrons.}
 \label{schema}
\end{figure}
Figure~\ref{schema} summarizes the scenario: In a thin layer of thickness $z_d$, determined by cascading and damping of the turbulence behind the forward shock, turbulence subjects electrons to stochastic re-acceleration. On account of the dominance of advection over diffusion, shock acceleration at $z=0$ provides accelerated electrons that are fed into the orange-shaded region of re-acceleration and leave it after time $t=z_d/v_\mathrm{adv}$. Electrons follow an $z$-$t$ characteristic, and therefore the entire spatial dependence of the electron density is given by the time evolution of the spectrum. The continuity equation then collapses to an initial-value problem of spectral evolution of $N(p,t)$, where 
\be
4\pi\,r^2\,N(r,p,t)=N(p,t=0)\,\delta\left(z-v_\mathrm{adv}\,t\right)\ ,
\label{eq:5}
\ee
and
\be
\frac{\partial N(p,t)}{\partial t}=\frac{\partial }{\partial p}\left(D_p(p)\,p^2\frac{\partial }{\partial p}\left(\frac{N(p,t)}{p^2}\right)\right) \ .
\label{eq:6}
\ee
Here, $D_p(p)$ is the diffusion coefficient in momentum space. For simplicity, we set the initial condition 
\be
N(p,t=0)=N_0\, p^{-2} \,\Theta\left(p-p_\mathrm{min}\right)\,\Theta\left(p_\mathrm{max}-p\right)\ .
\label{eq:7}
\ee
which corresponds to an unmodified strong shock in hydrogen gas \citep{1978MNRAS.182..147B} with cut offs at high and low energy to satisfy the Dirichlet boundary conditions at $p=0$ and $p=\infty$.

\subsection{Modeling}
If the momentum diffusion coefficient has the form $D_p(p)=D_0\,p^2$, then a complete analytical solution to Equation~\ref{eq:6} is known \citep{1962SvA.....6..317K}.
In general, the momentum diffusion coefficient has a more complex form and also implicitly depends on time through its spatial variation with $x(t)=v_\mathrm{adv}\,t$. 
We are not aware of any analytical solution for equation~(\ref{eq:6}) with arbitrary momentum diffusion coefficient $D_p(p)$, therefore it is solved numerically. In Section~\ref{diffacc} we established that only transit-time damping of fast-mode waves may be fast enough to modify particle spectra inside SNR. Under certain assumptions, among them isotropy of the particle distribution function, we found the acceleration time independent of momentum. Expecting that some of the assumptions break down for particles of higher energy, for the following discussion we shall therefore set: 
\be
D_p(p)=\frac{p^2}{\tau_\mathrm{acc}}f(p)\,,
\label{diff_coeff}
\ee
where $\tau_\mathrm{acc}$ is the acceleration time discussed in the previous section and $f(p)$ is a dimensionless function defined as:
\begin{displaymath}
f(p)=\begin{cases} 1 & \text{for\,\,}  p\le p_0\\ \big(\frac{p}{p_0}\big)^{-m} &\text{for\,\,}  p\ge p_0. \end{cases}
\end{displaymath}
Above $p_0$ the diffusion coefficient changes, where the power index $m$ determines how quickly the acceleration time increases with increasing particle energy.

The form of the momentum diffusion coefficient (Eq.~\ref{diff_coeff}) permits rewriting the reduced continuity equation~(\ref{eq:6}) in dimensionless coordinates. The acceleration time $\tau_\mathrm{acc}$ is the scale of a new dimensionless time coordinate $x$, and the new momentum coordinate $\tilde{p}$ is normalized with the critical momentum $p_0$ at which the behaviour of the diffusion coefficient changes,
\be
 x=\frac{t}{\tau_\mathrm{acc}}\qquad \tilde{p}=\frac{p}{p_0}
\ . 
\ee
Written in these dimensionless coordinates the continuity equation~(\ref{eq:6}) reads
\be
\frac{\partial N}{\partial x}=\frac{\partial }{\partial \tilde{p}}\left(f(\tilde{p})\tilde{p}^4\frac{\partial }{\partial \tilde{p}}\frac{N}{\tilde{p}^2}\right) \,
\label{Red_Eq}
\ee
and must be solved for $0\le x\le T=z_d/v_\mathrm{adv}/\tau_\mathrm{acc}$, where $T$ is the total available time in units of the acceleration time, previously estimated to be at most a few.

One immediately recognizes that the new equation depends on two parameters only. The parameter $x$ represents the relation between the age of the system and its acceleration efficiency. Second, the power index $m\in[0,1]$ of the momentum diffusion coefficient shapes particle spectra at $\tilde{p} > 1$. 

Note that the spatial and time coordinates are equivalent in our model. In the thin region of thickness $z_d$ behind the shock, where we expect strong turbulence, particle spectra will evolve as given in Equation~\ref{Red_Eq}. As $z_d$ may be only a few per cent of a light-year, and the projection of spherical shells on the sky plane will distribute its emission over a large area, the spatial variation of the radio spectrum will probably not be resolvable by current radio observatories. Hence, it may suffice to calculate the average spectrum with the region of strong turbulence (colored light red in Figure~\ref{schema}), for which we need to integrate the particle number density over time: 
\be
N_\mathrm{ave}(\tilde{p},T)_{}=\frac{1}{T} \int_{0}^{T}\,N(\tilde{p}, x)\,dx\,.
\label{TotNumD}
\ee
Once particles have left the region of turbulence, their spectra will evolve little, because re-acceleration is by definition inefficient and energy losses are slow. For most of the volume between the contact discontinuity and the forward shock, shaded yellow in Figure~\ref{schema}, we therefore expect the particle spectrum to be given by $N(\tilde{p}, x=T)$. 

In the next section we will present solutions $N_\mathrm{ave}(\tilde{p},T)$ for various $x$ and $m$. Additionally, we demonstrate how different initial spectra impact the solution. In fact, modified shocks generate softer spectrum at low energies than predicted by DSA. 


\section{Results}
\subsection{Particle spectra}
\begin{figure}[ht]
 \resizebox{\hsize}{!}{\includegraphics[angle=0]{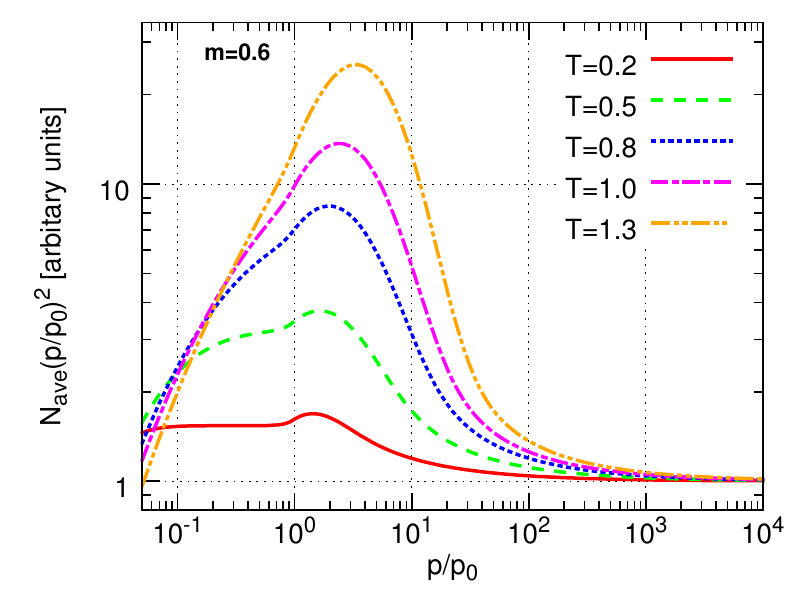}}
 \caption{Averaged electron number density, $N_\mathrm{ave}$, at different times $T$ for fixed index, $m=0.6$.}
 \label{age}
\end{figure}

In the following we set $N_0=1$ and initially $N(p,t=0)=N_0\,p^{-2}$. In Figure~\ref{age} we present the integrated particle number density, $N_\mathrm{ave}$, for different $T$ but fixed power index $m=0.6$ (cf. Equation~\ref{diff_coeff}). To be noted from the figure is the substantial flux enhancement near $p_0$ even if the available time is only a fraction of the acceleration time. This demonstrates that stochastic reacceleration can be important in SNR, despite the relatively long acceleration time. The assumed turn-over in the momentum dependence of the diffusion coefficient at $p/p_0=1$ causes the peak in $N_\mathrm{ave}\,p^2$ to be always close to $p_0$. 

\begin{figure}[ht]
 \resizebox{\hsize}{!}{\includegraphics[angle=0]{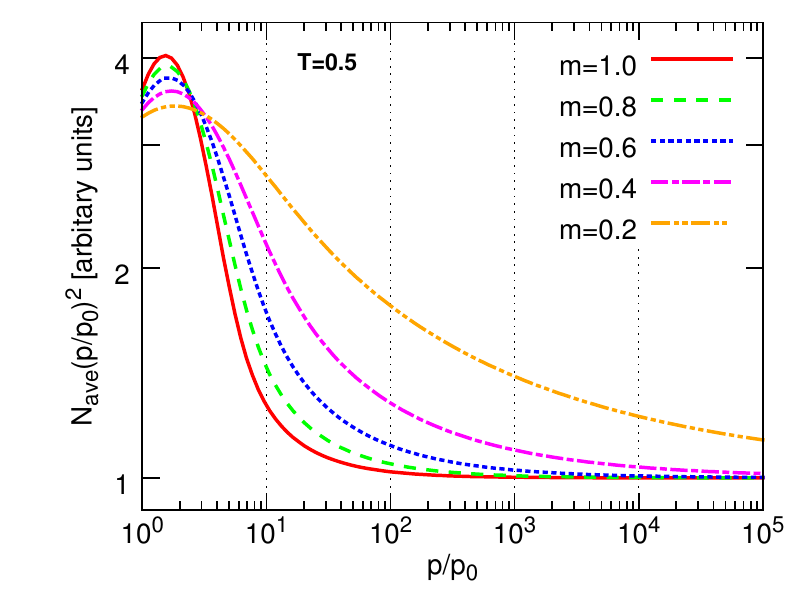}}
 \caption{Averaged electron number density, $N_\mathrm{ave}$, plotted for various power indices of the momentum diffusion coefficient, $m$, but fixed $T=0.5$.}
 \label{coeff}
\end{figure}

\begin{figure}[ht]
 \resizebox{\hsize}{!}{\includegraphics[angle=0]{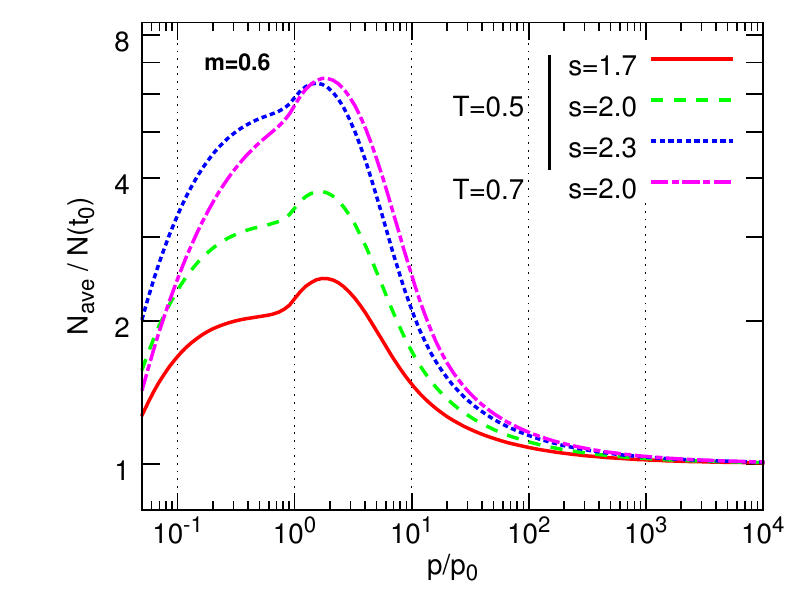}}
 \caption{Averaged electron number density, $N_\mathrm{ave}$, normalized by the initial distribution, $N(t_{0})=N_0 p^{-s}$, plotted for different initial indices, $s$.}
 \label{initC}
\end{figure}

The tail toward larger momenta is largely determined by the power index of the momentum diffusion coefficient, $m$. Figure~\ref{coeff} shows spectra for various $m$ and fixed time $T=0.5$. To be noted is that for small or moderate $m$ the spectral bump near $p_0$ has a high-energy tail that extends over more than 2 decades in momentum. To the outside observer that would appear as a softer spectrum at low momenta compared with that at very high momenta.
The question arises to what degree the spectral modifications imposed by stochastic reacceleration depend on the initial spectrum produced at the shock. Recall that for Figures~\ref{age} and \ref{coeff} we assumed $N(p,t=0)=N_0\,p^{-2}$, i.e. the test-particle solution for diffusive shock acceleration. Nonlinear modification of the shock and the properties of the scattering turbulence upstream of the forward shock can soften or harden the initial spectrum. To test whether the effect of stochastic reacceleration downstream is largely independent of the initial spectrum, we plot in figure~\ref{initC} the modification factor $N_\mathrm{ave}/N(p,t=0)$. For ease of exposition, the initial spectrum is assumed to follow a power law, $N(t_{0})=N_0 p^{-s}$, where we vary the index $s$. The form of the momentum diffusion coefficient is as in Equation~\ref{diff_coeff} with fixed $m$. We observe that the choice of initial spectral index determines mainly the amplitude of the spectral bump, whereas its shape is weakly affected. There is degeneracy between the parameters $m$ and $T$, visible, e.g., in the similarity of the spectral modification for $T=0.5$ and $s=2.3$ with that for $T=0.7$ and $s=2.0$. The initial conditions and the details of diffusive acceleration at the shock are largely irrelevant for the spectral characteristics provided by stochastic reacceleration in SNR.



\subsection{Radio synchrotron emission}
In strongly turbulent magnetic field with $\delta B\approx B_0$ the standard synchrotron emissivity is not applicable. In Appendix~\ref{a3} we have derived an analytical approximation to the synchrotron emissivity for a turbulent field with Gaussian distribution of amplitudes (cf. Equation~\ref{eq:a17}), that we shall use to calculate the radio spectral index. The main difference to the standard formula is a slower cut off $\propto \exp\left(-\nu^{2/3}\right)$.

\begin{figure}[ht]
 \resizebox{\hsize}{!}{\includegraphics[angle=0]{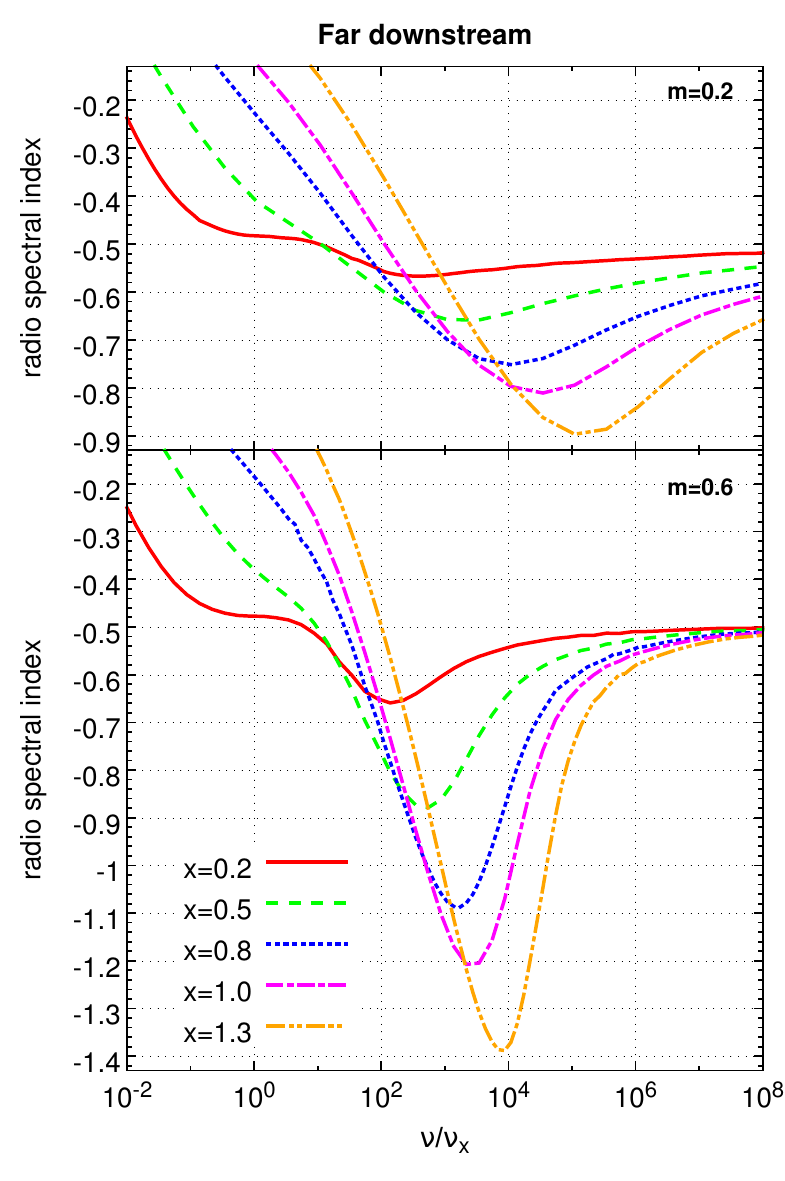}}
 \caption{Radio spectral index of the far downstream region of an SNR, plotted for different times $T$ and 2 choices of $m$.}
 \label{synch2}
\end{figure}

Having established that the choice of initial particle spectrum plays a minor role and can be compensated with adjustments in the dimensionless time, we calculate radio spectra only for $N(p,t=0)=N_0\,p^{-2}$, i.e. the radio spectral index at high frequencies is $\alpha=-0.5$.
As we use a dimensionless momentum coordinate, the synchrotron frequency is also dimensionless and normalized to the synchrotron frequency $\nu_x$ of electrons of momentum $p_0$ in a magnetic field of amplitude $B_\mathrm{rms}$,
\be
\nu_x=\nu_0 (B_\mathrm{rms},p_0)
=\frac{3\,e}{4\pi\,m_e^3\,c^3}\,B_\mathrm{rms}\,p_0^2\ .
\label{syncfreq}
\ee
The radio spectral index of the inner region, shaded yellow in Figure~\ref{schema}, must be calculated with $N(\tilde{p},x=T)$ and is shown in Figure~\ref{synch2}. Note that it is at the same time the radio spectrum at the inner edge of the region of reacceleration, and so it reflects the final state of the electron spectrum after experiencing momentum diffusion for a time $x=T=z_d/v_\mathrm{adv}/\tau_{acc}$.

\begin{figure}[ht]
 \resizebox{\hsize}{!}{\includegraphics[angle=0]{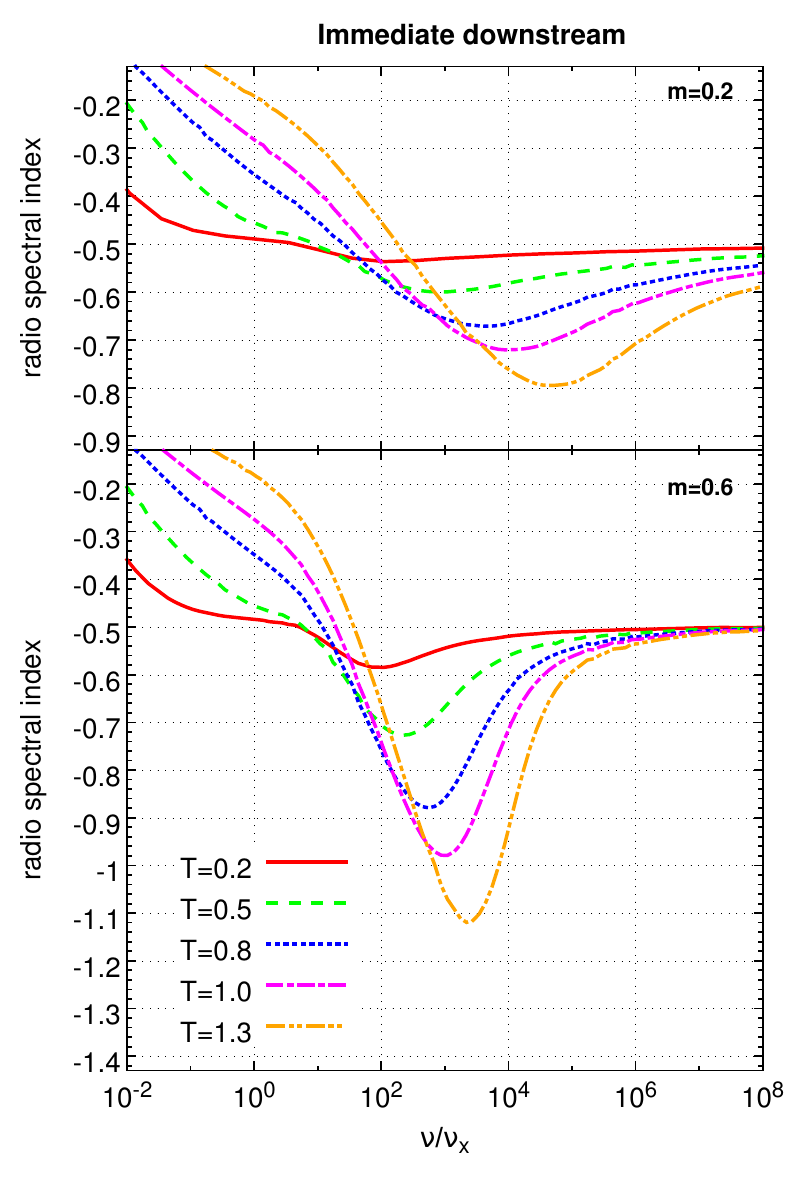}}
 \caption{Radio spectral index of the shell where reacceleration occurs, plotted for different times $T$ and 2 choices of $m$.}
 \label{synch}
\end{figure}

In Figure~\ref{synch} we present the radio spectral index, $\alpha$, of the region in which we expect substantial momentum diffusion (shaded red in Figure~\ref{schema}), i.e. computed using the average electron spectrum $N_\mathrm{ave}$. For ease of comparison, we chose the same time period, $T$, or thickness $z_d$, and 2 choices of momentum dependence of the acceleration time at high energy. 

We have seen in Figures~\ref{age}--\ref{initC} that a spectral bump results at a few $p_0$. Consequently, the radio spectra below approximately $10\,\nu_x$ are hard, above $10\,\nu_x$ they are soft, and eventually they approach those provided by the forward shock, here taken as $\alpha=-0.5$. The characteristics of the radio spectra are the following:
\begin{itemize}
\item The spectral modification in the far downstream region (cf. Figure~\ref{synch2}) is slightly stronger than that in the shell where momentum diffusion operates, because all electrons far downstream have sampled the full effect of reacceleration and have since experienced little change in energy, whereas the spectrum in the immediate downstream reflects an average of the spectral modification as it builds up. The total radio spectrum will be a mixture between the two.
\item Whenever the hardening at low frequencies below $\nu_x$ is significant, the spectral index changes rapidly with frequency, i.e. the spectral curvature is strong and should be detectable.
\item The softest spectra are observed at a few hundred to a thousand $\nu_x$. For soft radio spectra from SNR it is therefore sufficient, if $\nu_x\approx 10$~MHz, corresponding to $p_0\approx 150\ \mathrm{MeV}/c$ for $B_\mathrm{rms}\approx 25\ \mathrm{\mu G}$.
\item Less than one acceleration time is needed to soften the radio spectrum to $\alpha\simeq -0.65$. As the thickness of the acceleration region is $z_d=v_\mathrm{adv}\,T\,\tau_\mathrm{acc}$, for a reacceleration time of a few years and an advection speed of 1000~km/s we find that a thickness of $z_d\approx 3\cdot 10^{-3}$~pc is sufficient which in most cases is not resolvable.
\item If the increase of the reacceleration timescale with momentum is slow, i.e. $m$ is small, soft radio spectra with very little curvature can be maintained over 3 decades in frequency. In contrast, for $m=0.6$ spectral curvature is much stronger and should be detectable, in particular from the far downstream region.
\end{itemize}

\section{Summary and conclusions}
We have investigated the role of stochastic reacceleration in SNR with a view to probe whether or not it can account for the wide range of radio spectral indices observed among the more than 200 galactic SNR \citep{2009BASI...37...45G}. Turbulence that can change a particle's energy should be commonplace near the forward shocks of SNR. Cosmic-ray-driven instabilities operate in the upstream region. Other types of turbulence are excited at the shock itself. Whatever its nature, the turbulence will be advected to the downstream region where it has time to scatter energetic charged particles in pitch angle and momentum, until it is damped away. Reacceleration is therefore expected to be efficient, if anywhere, mostly in a potentially thin region behind the forward shock.

We calculated the momentum-diffusion coefficient for 3 types of turbulence, among which only transit-time damping of fast-mode waves operates on timescales of one or a few years. Incidentally, the energetic particles may be the dominant agent of damping for certain directions of wave propagation and thus harvest much of the turbulent energy. The acceleration time for transit-time damping of fast modes is found independent of energy, but expected to increase at higher energies on account of various inefficiencies. 

\mkp{In the case of small-scale non-resonant modes and the large-scale MHD turbulence arising from shock rippling, our estimates of the reacceleration rate are relatively simple. We feel that a more thorough treatment is not warranted on account of the long acceleration time that we derive. Transit-time damping of fast-mode waves is a much more promising process, for which we solve a resonance integral over the wave power spectrum. The main uncertainty here lies in the amplitude and spectral distribution of the waves, for which we here use generic arguments and cascading rates determined on the basis of detailed MHD simulations. While the wave power spectrum in a particular SNR will depend on the forward-shock speed and the properties of the upstream medium in that object, further work is needed to better understand the driving of fast-mode turbulence at astrophysical shocks with efficient particle acceleration.} 

Low-energy cosmic rays have a small mean free path, and so they are efficiently tied to the background plasma. Effectively, they advect on a characteristic in time $t$ and the spatial coordinate $z$ which describes the downstream distance to the forward shock. If reacceleration occurs only in a thin layer behind the shock, the cosmic-ray transport equation can be reduced to an initial-value problem, describing how the cosmic-ray spectrum is continuously deformed as the particles advect through the layer. Further inside, the particle spectrum is expected to change little and remain that calculated for the inner edge of the turbulence layer.

We numerically solved the reduced transport equation and found that cosmic-ray spectra develop a bump whose shape is largely independent of the initial spectrum assumed at the forward shock. To be noted is that the spectral bump can have an amplitude of a few hundred per cent, even if crossing the layer of efficient reacceleration takes less than one acceleration time. The shape of the high-energy tail of the bump depends on how quickly the acceleration time increases at high energies.

We calculated the synchrotron emissivity of electrons in a turbulent magnetic field with Gaussian distribution of amplitudes. Using that emissivity we determined the radio spectral index separately for the thin shell, where reacceleration occurs, and for the remaining interior of the SNR. The flux ratio between the two depends on the evolutionary history of the SNR and the supernova type, and so calculating total emission spectra can be done only once an object and its age are specified. Here, we only discuss the spectral index of radio emission from the two regions.

At low frequencies, where we observe particles subjected to reacceleration with energy-independent acceleration time, the radio synchrotron spectra tend to be hard with substantial curvature that should be evident in spectra covering 2 decades in frequency or more. Our results suggest that confusion with thermal or plerionic emission may be the culprit in cases of spectral indices close to $\alpha\approx 0$ extending over a large part of the radio band.

At higher frequencies, where we expect to see electrons that experienced momentum diffusion with an acceleration time increasing with particle energy, the radio synchrotron spectra are soft with indices $\alpha$ between $-0.6$ and $-0.7$ with little curvature for a slow increase of the diffusion coefficient with energy. About one acceleration time is sufficient to soften the radio spectrum by $\Delta \alpha \simeq -0.15$. The interiors of the remnants produce slightly softer radio spectra than does the shell where reacceleration occurs. Thus a modest reacceleration of electrons downstream of the forward shocks can explain the soft spectra observed from many galactic SNR.

\begin{acknowledgements}
We acknowledge support by the Helmholtz Alliance for Astroparticle Physics HAP
funded by the Initiative and Networking Fund of the Helmholtz Association. 
\end{acknowledgements}

\bibliographystyle{aa}
\bibliography{References}

\appendix
\section{Solving the resonance integral for fast-mode waves}
\subsection{Parallel-propagating waves}\label{a1}
For nearly parallel-propagating waves with $\mu_k=\cos\theta_k\simeq \pm 1$ we find for
the angle-dependent wavelength, $\lambda_\mathrm{c}$, at which the wave spectrum cuts off
\be
\lambda_\mathrm{c}=\lambda_\mathrm{max}\,\frac{c_\mathrm{s}^4\,(1-\mu_k^2)^2}{2.7\,V^4}\qquad\mathrm{and}\ \ 
\vert\mu_k\vert\ge\sqrt{1-\frac{2.7\,V^2}{c_\mathrm{s}^2}}\ .
\label{eq:a1a}
\ee
Note that there will be other damping mechanisms that will remove wave energy at some scale 
$\lambda_\mathrm{min}$, which we can treat as an upper limit to $\mu_k$ in the formula above.

If $\vert \mu_k\vert\simeq 1$, then resonance is achieved at $\mu\simeq 0$, and the term $c_s/(\mu_k\,c)$ in the argument of the exponential in Equation~\ref{eq:1.8} can be ignored.
A change of integration variable
\be
\mu\,\longrightarrow\,x=\frac{\mu}{\sqrt{1-\mu^2}}
\label{eq:a1}
\ee
turns Equation~\ref{eq:1.8} into 
\begin{align}
D_p\simeq&\frac{\sqrt{\pi}\,p^2\,c\,U_\mathrm{fm}}{8\,\epsilon\,B_0^2\,\beta\,\sqrt{\lambda_\mathrm{max}}}\, \int d\mu_k\ \frac{\vert \mu_k\vert}{\sqrt{\lambda_c}}\,\nonumber \\
&\cdot
\int_{-\infty}^\infty dx\ \left(1+x^2\right)^{-3}\,\exp\left[-\frac{x^2}{\epsilon^2}\right]\ .
\label{eq:a2}
\end{align}
The second integral in $x$ yields approximately $\epsilon$ for $\epsilon\simeq 0.5 --1$. To compute the first integral we insert $\lambda_c$ according to Equation~\ref{eq:a1a} and recall that there must be a limit $\lambda_\mathrm{min}$ to the turbulence spectrum, thus finding
\be
D_p\simeq \frac{\sqrt{2.7\,\pi}\,p^2\,c\,U_\mathrm{fm}}{16\,B_0^2\,\beta\,\lambda_\mathrm{max}}\,
\frac{V^2}{c_s^2}\,\ln\frac{\lambda_\mathrm{max}}{\lambda_\mathrm{min}}\ .
\label{eq:a3}
\ee
\subsection{Perpendicular-propagating waves}\label{a2}
For nearly perpendicular-propagating modes with $\vert\mu_k\vert\le 0.39$ damping by electrons dominates, and the cut off is found at
\begin{align}
\lambda_\mathrm{c}&=(4\cdot 10^{-6})\,\frac{\lambda_\mathrm{max}\,c_\mathrm{s}^4}{\mu_k^2\,V^4}\,
\exp\left(-\frac{1}{10\,\mu_k^2}\right)\ ,\nonumber \\
&\mathrm{where}\quad \left(\frac{\vert\mu_k\vert}{0.22}\right)^2\,
\left(2\,\ln\frac{c_\mathrm{s}}{14\,V}-\ln\frac{\vert\mu_k\vert}{0.39}\right) \le 1\ .
\label{eq:a4b}
\end{align}

\begin{figure}[ht]
 \resizebox{\hsize}{!}{\includegraphics[angle=0]{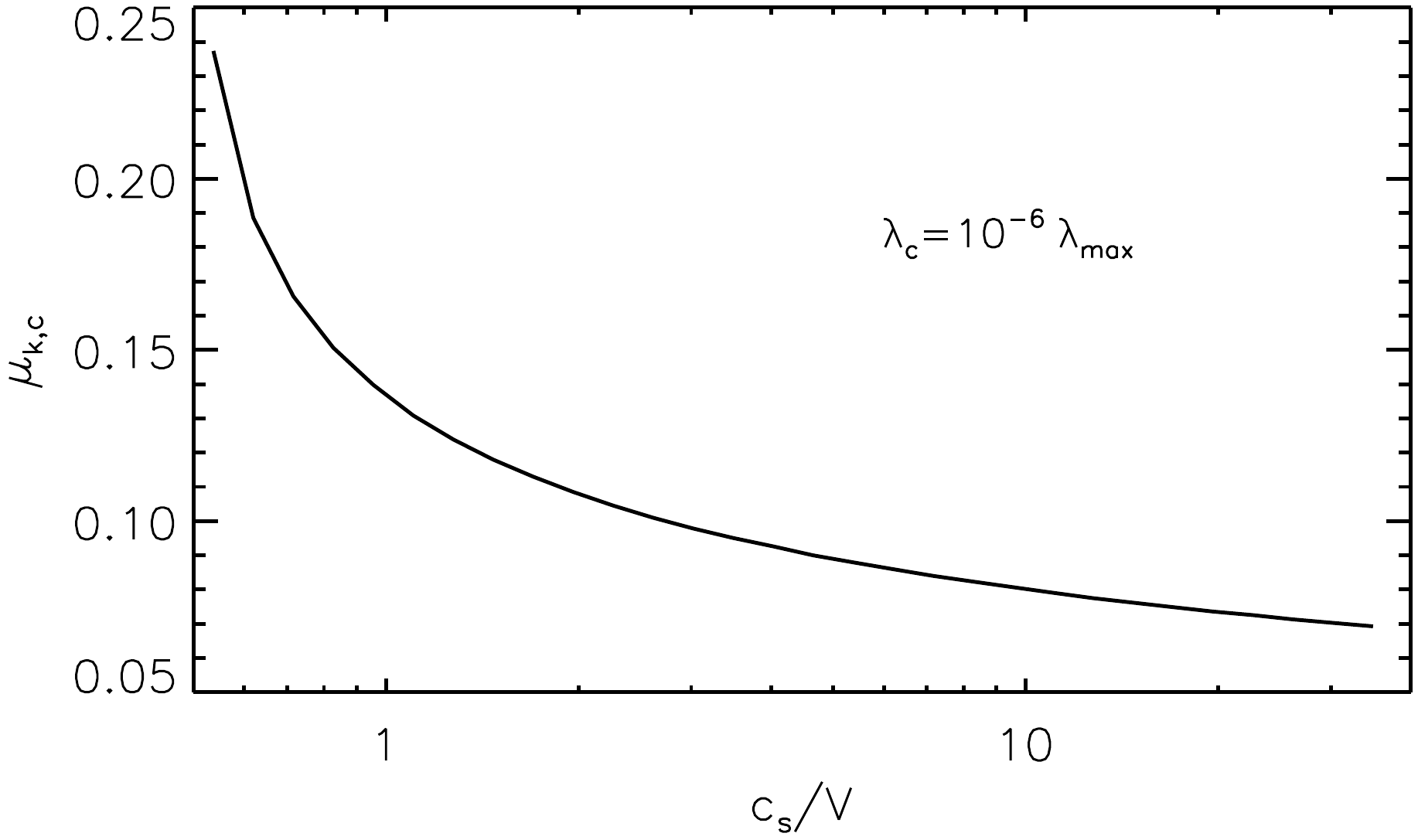}}
 \caption{The critical wave direction, $\mu_\mathrm{k,c}$, as function of the velocity amplitude at the driving scale, $V$, below which $\lambda_\mathrm{c}=10^{-6}\,\lambda_\mathrm{max}$.}
 \label{lambda}
\end{figure}
Equation~\ref{eq:a4b} suggests that the cut-off wavelength of cascading very rapidly falls off with decreasing $\mu_k$. Figure~\ref{lambda} displays the critical wave-angle cosine, $\mu_{k,c}$, at which $\lambda_\mathrm{c}=10^{-6}\,\lambda_\mathrm{max}$ is reached for all $\vert\mu_k\vert\le \mu_{k,c}$. To be noted is that $\mu_{k,c}\simeq 0.1$  for a wide range of velocity amplitudes, $V$, at the driving scale, $\lambda_\mathrm{max}$. For simplicity, we may therefore assume that $\lambda_\mathrm{c}=\lambda_\mathrm{min}=\mathrm{const.}$ for all $\vert\mu_k\vert\le \mu_{k,c}$. 

Then Equation~\ref{eq:1.8} can be written as
\begin{align}
D_p\simeq &\frac{\sqrt{\pi}\,p^2\,c\,U_\mathrm{fm}}{8\,\epsilon\,B_0^2\,\beta\,\sqrt{\lambda_\mathrm{max}\,\lambda_\mathrm{min}}}\,
\int d\mu\ \left(1-\mu^2\right)^{3/2}\nonumber \\
&\cdot \int_{-\mu_{k,c}}^{\mu_{k,c}} d\mu_k\ \vert \mu_k\vert\,
\exp\left[-\frac{(\mu-\frac{c_s}{c\,\mu_k})^2}{\epsilon^2\,(1-\mu^2)}\right]
\ .
\label{eq:a4}
\end{align}
Changing the variable of integration from $\mu_k$ to $y=c_s/(c\,\mu_k)$ simplifies this expression to 
\begin{align}
D_p\simeq &\frac{\sqrt{\pi}\,p^2\,c\,U_\mathrm{fm}}{4\,\epsilon\,B_0^2\,\beta\,\sqrt{\lambda_\mathrm{max}\,\lambda_\mathrm{min}}}\,\frac{c_s^2}{c^2}\,
\int d\mu\ \left(1-\mu^2\right)^{3/2}\nonumber \\
&\cdot \int_{\frac{c_s}{c\,\mu_{k,c}}}^\infty dy\ \frac{1}{y^3}\,
\exp\left[-\frac{(y-\mu)^2}{\epsilon^2\,(1-\mu^2)}\right]\ .
\label{eq:a5}
\end{align}
To be noted is that the lower limit of y-integration will be of the order $0.1$ for SNRs. Given the form of the integrand in the second integral, small $y$ will dominate the integral. Swapping the order of integration, approximating $y\ll 1$ in the argument of the Gaussian, and reusing the variable transformation in Equation~\ref{eq:a1}, we find with reasonable accuracy
\begin{align}
D_p\approx &\frac{\sqrt{\pi}\,p^2\,c\,U_\mathrm{fm}}{4\,\epsilon\,B_0^2\,\beta\,\sqrt{\lambda_\mathrm{max}\,\lambda_\mathrm{min}}}\,\frac{c_s^2}{c^2}\,\int_{\frac{c_s}{c\,\mu_{k,c}}}^\infty dy\ \frac{1}{y^3}\,
\nonumber \\
&\cdot \int_{-\infty}^\infty dx\ \frac{1}{\left(1+x^2\right)^{3}}\,
\exp\left[-\frac{x^2}{\epsilon^2}\right]\ .
\label{eq:a6}
\end{align}
Both integrals can be easily solved separately to yield the final expression for the momentum diffusion coefficient. We already noted in Appendix~\ref{a1} that the second integral is approximately $\epsilon$. 
\be
D_p\approx \frac{\sqrt{\pi}\,p^2\,c\,U_\mathrm{fm}}{8\,B_0^2\,\beta\,\sqrt{\lambda_\mathrm{max}\,\lambda_\mathrm{min}}}\,\mu_{k,c}^2\ .
\label{eq:a7}
\ee
which may be more conveniently written using the thermal energy density in the downstream plasma, $U_\mathrm{th}\simeq \rho\,c_s^2$, 
\be
D_p\approx \frac{p^2\,c}{40\,\sqrt{\pi}\,\sqrt{\lambda_\mathrm{max}\,\lambda_\mathrm{min}}}\,\frac{U_\mathrm{fm}}{U_\mathrm{th}}\,\mu_{k,c}^2\ .
\label{eq:a8}
\ee
\section{Synchrotron emissivity for turbulent magnetic field}\label{a3}
For isotropic magnetic turbulence we may start with the angle-averaged spectral power per electron in a magnetic field of constant amplitude, which is well approximated with
\citep{1986A&A...164L..16C}
\be
P_\nu =C\,B\,\left(\frac{\nu}{\nu_0}\right)^{1/3} 
\exp\left(-\frac{\nu}{\nu_0}\right)\ ,
\label{eq:a9}
\ee
where
\be
C=1.8\,\frac{\sqrt{3}\,e^3}{4\pi\,m_e\,c^2}\quad\mathrm{and}\quad \nu_0=\frac{3\,e}{4\pi\,m_e\,c}\,B\,\gamma^2
\ .
\label{eq:a10}
\ee
We observe an individual electron radiating for only the Larmor period of a non-relativistic electron, which is considerably shorter than the period of the MHD waves that comprise the magnetic turbulence in SNRs. The instantaneous contribution to the synchrotron emissivity of an individual electron is therefore well described by a constant local magnetic acceleration, followed by averaging over all possible local magnetic-field strengths.  

We suppose the magnetic-field amplitude follows a Gaussian probability distribution,
\be
P_B = \frac{\sqrt{2}}{\sqrt{\pi}\,B_\mathrm{rms}}\,
\exp\left(-\frac{B^2}{2\, B_\mathrm{rms}^2}\right)
\ .
\label{eq:a11}
\ee
The effective spectral power therefore is
\be
P_\mathrm{eff} = \int_{0}^\infty dB\ P_\nu P_B
\ .
\label{eq:a12}
\ee
Denoting $x=B/B_\mathrm{rms}$ and $\nu_c=\nu_0(B_\mathrm{rms})$ we find
\begin{align}
P_\mathrm{eff} =& \sqrt{\frac{2}{\pi}}\,C\,B_\mathrm{rms}\, \left(\frac{\nu}{\nu_c}\right)^{1/3} \nonumber \\
&\cdot\ \int_{0}^\infty dx\ x^{2/3}\, \exp\left(-\frac{x^2}{2}-\frac{\nu}{\nu_c\, x}\right)
\ .
\label{eq:a13}
\end{align}
For $\nu\ll \nu_c$ the last exponential is irrelevant and the integral yields $1$. The low-frequency spectral power therefore is 
\be
P_\mathrm{eff} \simeq\sqrt{\frac{2}{\pi}}\,C\,B_\mathrm{rms}\, \left(\frac{\nu}{\nu_c}\right)^{1/3} \qquad \mathrm{for}\ \nu\ll \nu_c\ .
\label{eq:a13a}
\ee
For high frequencies, $\nu\gg \nu_c$, we transform to $y=x^{5/3}$, implying $dy=(5/3)\, x^{2/3}\, dx$. Then
\begin{align}
P_\mathrm{eff} = &\frac{6\,C\,B_\mathrm{rms}}{5\,\sqrt{2\pi}}\, \left(\frac{\nu}{\nu_c}\right)^{1/3} \nonumber \\
&\cdot\ 
\int_{0}^\infty dy \ \exp\left(-\frac{y^{6/5}}{2}-\frac{\nu}{\nu_c\, y^{3/5}}\right)
\ .
\label{eq:a14}
\end{align}
We use the method of steepest ascent to solve the integral \citep[e.g.][]{wong..asymptotic}. 
The negative argument of the exponential, $\exp\left[-F(y)\right]$, is Taylor-expanded 
around its minimum at $y_0=(\nu/\nu_c)^{5/9}$, giving
\be
F(y)\simeq \frac{3}{2}\,\left(\frac{\nu}{\nu_c}\right)^{2/3}
+\frac{(y-y_0)^2}{2}\,\frac{27}{25}\,\left(\frac{\nu}{\nu_c}\right)^{-4/9}
\ .
\label{eq:a15}
\ee
The integral in Equation~\ref{eq:a14} then reduces to a Gaussian and yields
\begin{align}
P_\mathrm{eff} \simeq&\frac{C\,B_\mathrm{rms}}{\sqrt{3}}\, \left(\frac{\nu}{\nu_c}\right)^{5/9}\,
\exp\left(-\frac{3}{2}\,\left(\frac{\nu}{\nu_c}\right)^{2/3}\right)\nonumber \\
&\cdot \left(1+\mathrm{erf}\left[\sqrt{\frac{27}{50}}\, 
\left(\frac{\nu}{\nu_c}\right)^{1/3}\right]\right)
\ .
\label{eq:a16}
\end{align}
\begin{figure}[ht]
 \resizebox{\hsize}{!}{\includegraphics[angle=0]{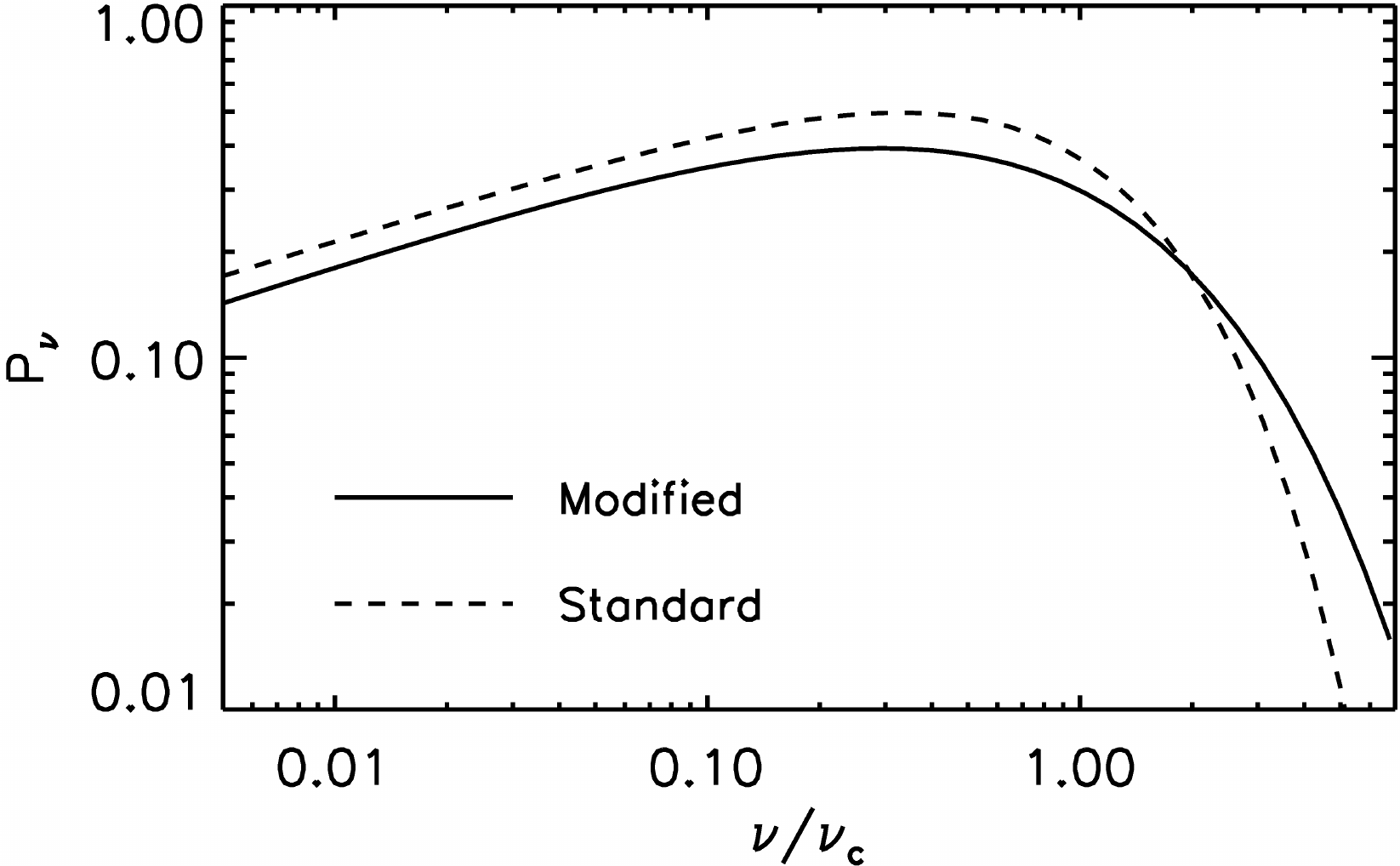}}
 \caption{Comparison of the standard expression for the spectral synchrotron power of electrons with that derived here for turbulent magnetic field with Gaussian distribution.}
 \label{sync}
\end{figure}
This is the high-frequency solution to the integral in Equation~\ref{eq:a13} which we need to combine with the low-frequency solution given in equation~\ref{eq:a13a}. Noting that the argument of the error function is slowly varying, we can replace the error function with a constant that is appropriate for frequencies slightly above $\nu_0$. We know that the normalization of the spectral power has to match that for a homogeneous magnetic field with $B=B_\mathrm{rms}$, because the energy loss rate is quadratic in $B$. Using an algebraic transition that is accurate in the normalization to within 1\% and matches the asymptotic behaviour, we finally obtain
\begin{align}
P_\mathrm{eff} \simeq &C\,B_\mathrm{rms}\,\sqrt{\frac{2}{\pi}}\, \left(\frac{\nu}{\nu_c}\right)^{1/3} \,
\exp\left(-\frac{3}{2}\,\left(\frac{\nu}{\nu_c}\right)^{2/3}\right)\nonumber \\
&\cdot \left(1+1.65\,\left(\frac{\nu}{\nu_c}\right)^{0.42}\right)^{0.53}
\ .
\label{eq:a17}
\end{align}
A comparison with the standard formula (Equation~\ref{eq:a9}) is given in Figure~\ref{sync}.
A corresponding formula for an exponential distribution of magnetic-field amplitudes can be found in \citet{2010ApJ...708..965Z}.
\end{document}